\documentclass[twocolumn]{emulateapj}

\lefthead{Law et al. 2009}
\righthead{Evidence for a Triaxial Dark Matter Halo}

\slugcomment{DRAFT: \today}

\begin{document}

\newcommand{\msun}{\ensuremath{\rm M_\odot}}
\newcommand{\msunyr}{\ensuremath{\rm M_{\odot}\;{\rm yr}^{-1}}}\newcommand{\Ha}{\ensuremath{\rm H\alpha}}
\newcommand{\Hb}{\ensuremath{\rm H\beta}}
\newcommand{\lya}{\ensuremath{\rm Ly\alpha}}
\newcommand{\Ntwo}{[\ion{N}{2}]}
\newcommand{\kms}{km~s\ensuremath{^{-1}\,}}
\newcommand{\ztwo}{\ensuremath{z\sim2}}
\newcommand{\zthree}{\ensuremath{z\sim3}}
\newcommand{\feh}{[Fe/H]}

\newcommand{\hst}{{\it HST}-ACS}

\title{Evidence for a Triaxial Milky Way Dark Matter Halo from the Sagittarius Stellar Tidal Stream}
\author{David R.~Law\altaffilmark{1,2}, Steven R. Majewski\altaffilmark{3}, Kathryn V. Johnston\altaffilmark{4}}

\altaffiltext{1}{Hubble Fellow.}
\altaffiltext{2}{Department of Physics and Astronomy, University of California, Los Angeles, CA 90095;
drlaw@astro.ucla.edu}
\altaffiltext{3}{Dept. of Astronomy, University of Virginia,
Charlottesville, VA 22904-0818 (srm4n@virginia.edu)}
\altaffiltext{4}{Department of Astronomy, Columbia University, New York, NY 10027 (kvj@astro.columbia.edu)}

\begin{abstract}

Observations of the lengthy tidal streams produced by  the destruction of the Sagittarius dwarf spheroidal (Sgr dSph)
are capable of providing strong constraints on the shape of the Galactic gravitational potential.
However, previous work, based on
modeling different stream properties in axisymmetric Galactic models 
has yielded conflicting results: while the angular precession of the Sgr leading arm is most consistent with a spherical or slightly oblate halo, the radial velocities of stars in this arm are only reproduced
by prolate halo models.
We demonstrate that this apparent paradox can be resolved by instead adopting a triaxial potential. 
Our new Galactic halo model, which simultaneously fits all well-established phase space constraints 
from the Sgr stream, provides
the first conclusive evidence for, and tentative measurement of, triaxiality in an individual dark matter halo.  
The Milky Way halo
within $\sim 60$ kpc is best characterized by a minor/major axis ratio of the isovelocity contours
 $c/a \approx 0.67$, intermediate/major axis ratio $b/a \approx 0.83$, and triaxiality parameter $T \sim 0.56$.
In this model, the minor axis of the dark halo is coincident with the 
Galactic $X$ axis connecting the Sun and the Galactic Center to within $\sim 15^{\circ}$, while the major
axis also lies in the Galactic plane, approximately along the Galactic $Y$ axis.

\end{abstract}


\section{INTRODUCTION}

One of the general predictions of structure formation within the prevailing cold dark matter (CDM) paradigm is that galaxy-scale dark matter haloes should be 
described by a triaxial density ellipsoid (e.g., Jing \& Suto 2002; Bailin \& Steinmetz 2005; Allgood et al. 2006, and references therein).
The characteristic axial ratios of these haloes are typically expected to be far from spherical (see, e.g., Bullock et al. 2002; Kuhlen et al. 2007; and references therein)
with characteristic central minor/major axis ratios (in the isovelocity contours) $c/a \sim 0.72$ and intermediate/major axis ratios $b/a \sim 0.78$ (Hayashi et al. 2007).
Adopting the triaxiality parameter of Franx et al. (1991),
\begin{equation}
T = \frac{1- b^2/a^2}{1- c^2/a^2}
\end{equation}
this corresponds to a typical value of $T = 0.81$.  As discussed by Hayashi et al. (2007), there can be significant variation of the axis ratios with radius
and between realizations of similar mass haloes: individual values for $c/a$ range from $\sim 0.6 - 0.9$ while $b/a$ can take values $\sim 0.6 - 1.0$.
Despite the near-ubiquitous predictions of triaxiality from such simulations, there has hitherto been little observational evidence to confirm triaxiality in specific individual
galaxies: while gravitational lensing (e.g., Hoekstra et al. 2004; Mandelbaum et al. 2006; Evans et al. 2009) and X-ray observations (e.g., Pointecouteau et al. 2005)
have indicated that non-sphericity appears to be common in
dark matter haloes, such studies are sensitive only to the integral of the density profile along the line of sight
and do not provide fully 3-D information.


The Milky Way Galaxy provides perhaps the best laboratory for testing predictions of halo sphericity 
since the tidal stream remnants of dwarf satellites orbiting in the Galactic halo
can be traced in three dimensions and 
provide sensitive probes of the underlying mass distribution.  The tidal tails emanating from the Sagittarius dwarf spheroidal galaxy 
(Sgr dSph)\footnote{We refer the non-expert reader to LJM05 (see particularly their Fig. 1) for an 
extended introduction to the general characteristics of the Sgr dwarf system.} 
have been used for such efforts since
shortly after the discovery of the dwarf by Ibata et al. (1994).  
Early efforts to model the MW --- Sgr system (e.g., Johnston et al. 1995, 1999; Velazquez \& White 1995; 
Edelsohn \& Elmegreen 1997; Ibata et al. 1997; 
G{\'o}mez-Flechoso et al. 1999; Helmi \& White 2001) generally concentrated on fitting the orbit of Sgr within an adopted Milky Way potential.
In more recent years however, compelling wide-field views of the extensive tidal streams associated with the dwarf have been
provided by the Two Micron All-Sky Survey (2MASS; Majewski et al. 2003) and Sloan Digital Sky Survey (SDSS; Belokorov et al. 2006),
and permitted much more detailed exploration of the underlying shape of the Galactic potential to be undertaken by
Helmi (2004), Mart{\'{\i}}nez-Delgado et al. (2004, 2007), Law, Johnston, \& Majewski (2005; hereafter LJM05), Fellhauer et al. (2006), and Cole et al. (2008).
Despite the prediction from CDM theory that the Milky Way should be best described by a triaxial Galactic potential, the majority of models to date 
(although cf. Gnedin et al. 2005)
have adopted an axisymmetric framework
 in which flattening is introduced only into the Galactic $Z$ axis (i.e. perpendicular to the Galactic disk).

While these recent simulations agree on many aspects of the Sgr orbit, no one model has
yet been capable of reproducing all of the observational data and these models reach different conclusions
about the shape of the Galactic halo depending on which observational data are weighted as primary constraints:
Some studies favor a 
mildly oblate halo (e.g., Johnston et al. 2005 [hereafter JLM05]; Mart{\'{\i}}nez-Delgado et al. 2007),
some an approximately spherical halo (e.g., Ibata et al. 2001; Fellhauer et al. 2006), 
and others a prolate halo (e.g., Helmi 2004).
The crux of this ``halo conundrum''  (highlighted by LJM05, and also discussed by Fellhauer et al. 2006, Mart{\'{\i}}nez-Delgado et al. 2007, Newberg et al. 2007, and Yanny et al. 2009)
is that in an axisymmetric Galactic potential it is not possible to simultaneously fit both the angular precession and distance/apparent radial velocity of stars in
the leading Sgr stream as it arcs through the North Galactic Cap towards the Galactic anticenter.
In order to match the angular coordinates of stars in the Sgr leading arm JLM05 and Fellhauer et al. (2006) required mildly oblate or nearly spherical models
(with $Z$-axis flattening $q_z \sim 0.9-1.05$ in the contours of the gravitational potential) for the Galactic dark halo.  
Such models cause the leading arm to reenter the Galactic disk in the vicinity of the Sun (to within $\sim 5-10$ kpc), in contradiction with
photometric distance estimates (e.g., Newberg et al. 2007) from SDSS.  Such a model 
also predicts a significant projection of the orbital velocities of leading tidal debris onto the observed line of sight, in contrast
to spectroscopic radial velocity data (presented by Law et al. 2004, LJM05, and more recently confirmed by Yanny et al. 2009).
In order to reproduce the radial velocity and distance trends  the halo must be prolate with  $q_z \sim 1.25$ (as demonstrated by
Helmi et al. 2004 and confirmed by LJM05), but this in turn cannot account for the observed  precession experienced by the tidal debris.

In this Letter we demonstrate that this ``halo conundrum'' is a consequence of the near-ubiquitous adoption of axisymmetric models for the Galactic
potential.  By using fully  triaxial models, similar to those predicted by standard CDM theory,
it is possible to simultaneously satisfy all constraints imposed by the structure of Sgr debris on the Milky Way's mass distribution.

\section{MODEL}
\label{model.sec}

We adopt a basic  formalism similar to that described by LJM05 (and described in greater detail by Law et al., {\it in prep.}) in which
the Milky Way is described by a smooth fixed gravitational potential
consisting of a Miyamoto-Nagai (1975) disk, Hernquist spheroid, and a logarithmic halo.
The respective contribution of these components to the gravitational potential is given by:

\begin{equation}
        \Phi_{\rm disk}=- \alpha {GM_{\rm disk} \over
                 \sqrt{R^{2}+(a+\sqrt{z^{2}+b^{2}})^{2}}},
\label{diskeqn}
\end{equation}
\begin{equation}
        \Phi_{\rm sphere}=-{GM_{\rm sphere} \over r+c},
\label{bulgeqn}
\end{equation}
\begin{equation}
        \Phi_{\rm halo}=v_{\rm halo}^2 \ln (C_1 x^2 + C_2 y^2 +C_3 x y + (z/q_z)^2 + r_{\rm halo}^2)
\label{haloeqn}
\end{equation}
where the various constants $C_1$, $C_2$, $C_3$ are given by
\begin{equation}
C_1 = \left(\frac{\textrm{cos}^2 \phi}{q_{1}^2} + \frac{\textrm{sin}^2 \phi}{q_{2}^2}\right)
\end{equation}
\begin{equation}
C_2 = \left(\frac{\textrm{cos}^2 \phi}{q_{2}^2} + \frac{\textrm{sin}^2 \phi}{q_{1}^2}\right)
\end{equation}
\begin{equation}
C_3 = 2 \, \textrm{sin}\phi \, \textrm{cos} \phi \left( \frac{1}{q_{1}^2} - \frac{1}{q_{2}^2}\right)
\end{equation}
This form for the halo potential permits flattening to be introduced along the three axes $q_1$, $q_2$, $q_z$, where $q_z$ represents the flattening 
perpendicular to the Galactic Plane, while $q_1$ and $q_2$ are free to rotate in the Galactic Plane at an angle $\phi$ to a right-handed Galactocentric $X,Y$
coordinate system\footnote{I.e., when $\phi = 0^{\circ}$, $q_1$ is aligned with the Galactic $X$ axis and
Eqn. \ref{haloeqn} reduces to $\Phi_{\rm halo}=v_{\rm halo}^2 \ln ([x/q_1]^{2}+[y/q_2]^2+[z/q_z]^2+r_{\rm halo}^{2})$.}.
Since it is only the {\it ratios} between these $q$ that have physical significance, we fix
$q_2 = 1.0$ and explore the effects of varying $q_1$ and $q_z$ in the range $1.0 - 1.8$, with $\phi = 0^{\circ} - 180^{\circ}$.
We do not consider the range $q_z, q_1 > 1.8$ since our formulation of the Galactic gravitational potential rapidly becomes unphysical for larger values.

We note that this method introduces the flattening directly into the gravitational potential (i.e. the isovelocity contours) for ease of calculation
and consistency with previous studies (e.g., LJM05).  We have also considered more physical models
in which axial flattenings are introduced into the density profile of 
an NFW (Navarro et al. 1996) halo model, and the resulting accelerations computed using the approximate form of the potential given by Lee \& Suto (2003).
Such models 
give slightly different orbital
paths, but the qualitative results discussed below remain essentially unchanged.


We assume that the Sun is located 8.0 kpc from the Galactic Center
and 28 kpc from the Sgr core (Siegel et al. 2007).  
We fix various constants in Equations \ref{diskeqn} - \ref{haloeqn} based on previous work by LJM05, adopting
$M_{\rm disk} = 1.0 \times 10^{11} M_{\odot}$, $M_{\rm sphere} = 3.4 \times 10^{10} M_{\odot}$, $a = 6.5$ kpc, $b = 0.26$ kpc, $c = 0.7$ kpc, $r_{\rm halo} = 12$ kpc.
Similarly, the position, radial velocity, and  instantaneous orbital plane of the Sgr dwarf are fixed as discussed in LJM05 (see also Law et al., in prep).

For each combination of $q_1$, $q_z$, and $\phi$ tested we proceed as follows;
(1) We fixed $v_{\rm halo}$ in Equation \ref{haloeqn} so that the LSR has an orbital velocity $v_{\rm LSR} = 220$ km s$^{-1}$.
(2) While the dynamics of the
{\it leading} arm of Sgr tidal debris are strongly dependent on the choice of Galactic potential, the {\it trailing} arm experiences very little precession
and its characteristic angular position (Majewski et al. 2003) and radial velocities (Majewski et al. 2004) can be well reproduced in any realistic model of the Galactic potential
given suitable choice of the  speed $v_{\rm tan}$ of Sgr perpendicular to the line of sight (as demonstrated by LJM05).  We therefore adjusted $v_{\rm tan}$  until the best match to the
trailing debris data was obtained.
(3) Massless test-particles were integrated along the orbit, and the quality of fit $\chi$ to the observational data calculated.

The accuracy of such test-particle orbits will necessarily be limited
since actual tidal debris (with a  range of orbital energies and angular momenta)
from massive satellites will deviate slightly from these orbits; leading/trailing arms will fall inside/outside the orbital path respectively (see, e.g., discussion by Johnston et al.
1995, 1999; Choi et al. 2007; Eyre \& Binney 2009).
However, previous work (e.g., LJM05) has demonstrated that for Sgr the test-particle orbits provide a good indication of the general trend of tidal debris (particularly for the observational
constraints upon which we focus) and suffice for the first-order estimate of the underlying gravitational potential presented here.

\section{RESULTS}

\begin{figure*}
\epsscale{1.0}
\plotone{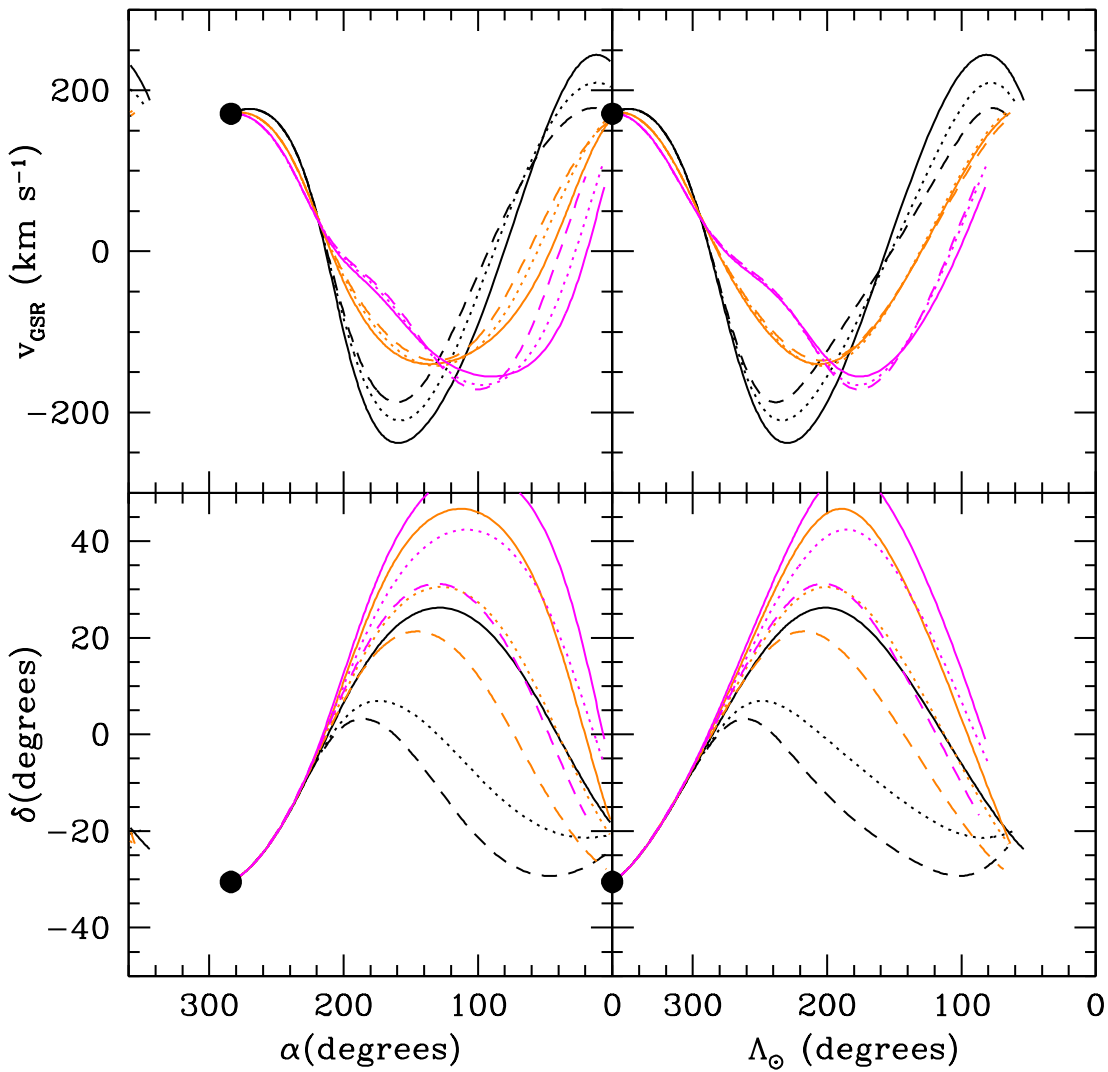}
\epsscale{1.0}
\caption{Model orbits are plotted 1.2 Gyr in advance of the current position of the Sgr dwarf (denoted by filled circles) illustrating the trend of declination ($\delta$; bottom panels)
and apparent radial velocity ($v_{\rm GSR}$; top panels) expected for leading arm tidal debris.
Black/orange/magenta curves respectively denote haloes in which $q_z = 1.0/1.3/1.6$ while solid/dotted/dashed curves correspond to values $q_y = 1.0/1.3/1.6$ (i.e. $\phi = 90^{\circ}$) respectively.
Given the split in the literature between describing position along the orbit in terms of the right
ascension ($\alpha$) or the more natural orbital longitude ($\Lambda_{\odot}$; see Majewski et al. 2003) we show these relations against both $\alpha$ and $\Lambda_{\odot}$
in left/right-hand panels respectively to aid comparison to plots published by previous authors.}
\label{OrbitCompare.fig}
\end{figure*}

We first illustrate the spirit of our results with the specific case of $\phi = 90^{\circ}$ (i.e. $q_1 = q_y$).
In Figure \ref{OrbitCompare.fig} we plot radial velocity (with respect to the Galactic Standard of Rest [GSR]) and declination along the section of orbit leading the Sgr dwarf for models
in which the Milky Way is taken to have various values of $q_y$ and $q_z$.  
While the trend of radial velocity is strongly affected by the choice of $q_z$ (i.e. different colors
in upper panels of Fig. \ref{OrbitCompare.fig}), $q_y$ has a comparatively minor effect on the velocities (difference between solid/dotted/dashed lines of a given color) 
in the key range $\alpha \sim 250^{\circ} - 100^{\circ}$.
In contrast, both $q_y$ and $q_z$ affect the precession of the leading arm 
as characterized by its declination $\delta$ for a given $\alpha$ or $\Lambda_{\odot}$ coordinate. 
As shown in the lower panels of Figure \ref{OrbitCompare.fig},
increasing $q_z$ shifts the arm to greater $\delta$ for fixed $q_y$ (i.e. changing from black, to orange, to magenta curves for a given line type in Fig. \ref{OrbitCompare.fig}),
and increasing $q_y$ decreases $\delta$ for fixed $q_z$ (i.e. changing from solid, to dotted, to dashed curves for a given line color in Fig. \ref{OrbitCompare.fig}).
These trends indicate that while both $q_y$ and $q_z$ govern the angular precession of the stream, it is primarily $q_z$ that affects the distance to the leading arm, 
altering the projection of the orbital velocity onto the line of sight.

\begin{figure*}
\epsscale{1.0}
\plotone{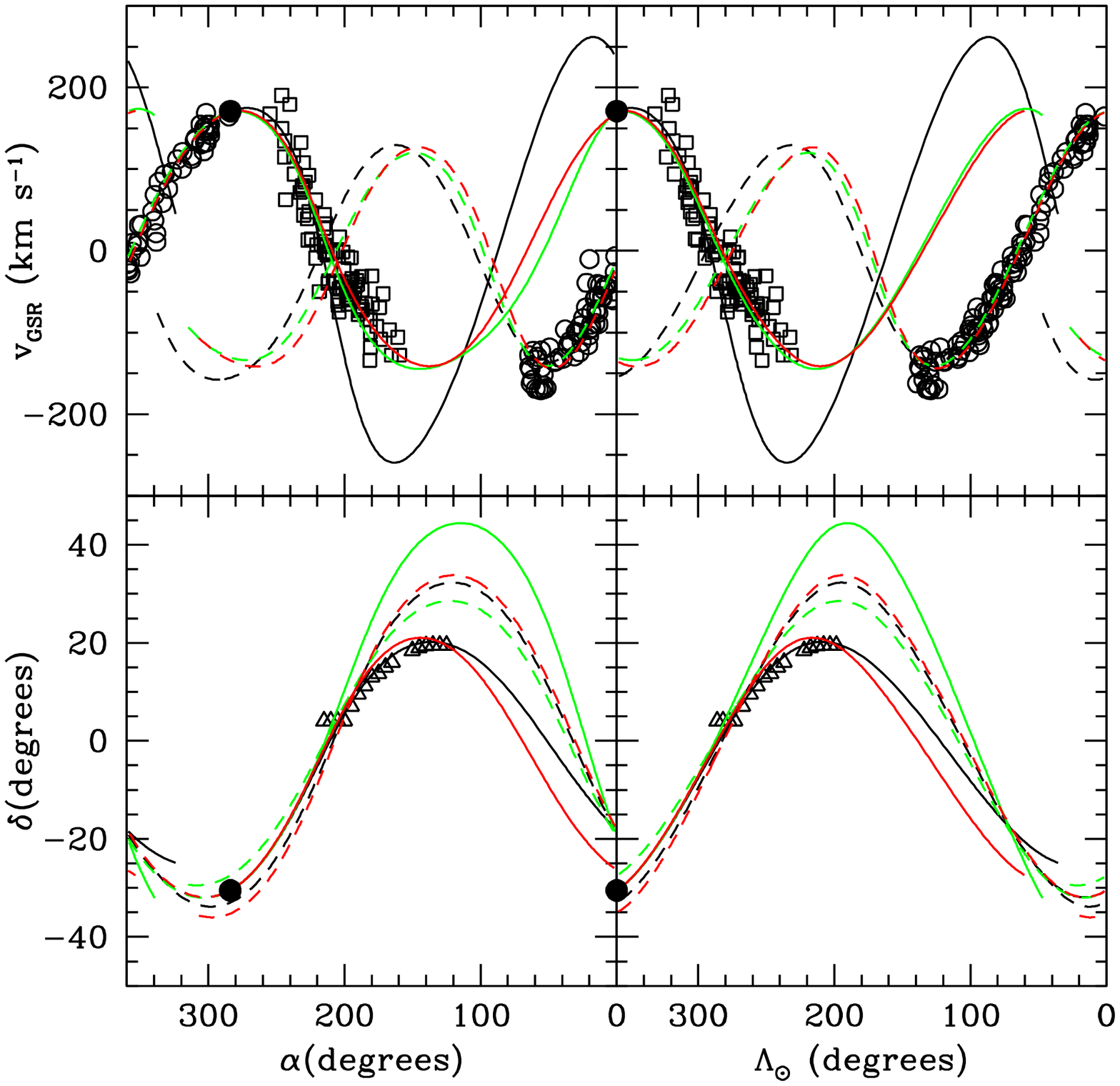}
\epsscale{1.0}
\caption{Model orbits  are plotted for three models: Axisymmetric halo best fitting leading arm precession (black curve), axisymmetric halo best fitting leading arm velocities (green curve),
and triaxial halo model (red curve).  Solid/dashed lines indicate the orbital path of Sgr leading/trailing its current position for 1.2 Gyr.
Open triangles represent SDSS data from Belokorov et al. (2006), open squares/circles represent leading/trailing arm M-giant radial velocity data
from Law et al. (2004) and Majewski et al. (2004) respectively.  Note that although the solid black curve fits the angular coordinates well it provides a poor match
to the radial velocities; the solid green curve provides a good match to the radial velocities but a poor match to the angular
coordinates; but the red curve matches both velocities and angular coordinates well.}
\label{ObsData.fig}
\end{figure*}

In the special case of $\phi = 90^{\circ}$, $q_z$ and $q_1$ are clearly separable parameters which may be constrained in turn.
In general however, $q_z$ and $q_1$ are not separable and must be constrained in tandem so that the resulting tidal stream
matches both the apparent radial velocity and the observed angular precession.
In Figure \ref{ObsData.fig} we summarize the three most relevant observational constraints:
trailing arm velocity data from Majewski et al. (2004; open circles), leading arm angular coordinates from Belokurov et al. (2006; open trianges),
and leading arm radial velocities from Law et al. (2004; see also LJM05 and Majewski et al. {\it in prep.}; open boxes).
While there is evidence for a ``bifurcation'' in the Sgr leading stream (Belokurov et al. 2006), 
we assume that the centroid of the stream is represented by the main, highest surface-brightness `southern' branch (see discussion by Yanny et al. 2009; Law et al. in prep).
As discussed in \S \ref{model.sec}, it is possible to fit the trailing arm velocity data in most reasonable models of the Galactic potential with suitable choice of the velocity
of Sgr along its orbit, and all of our models are designed to do so.

We define a $\chi^2$ statistic characterizing the quality of fit of the Sgr orbit in our various models of the Galactic halo to the angular coordinates and velocity data:
\begin{eqnarray}
\chi^2 =  \frac{1}{n_{v,{\rm lead}} - 3} \sum_i \frac{(v_{\rm orbit} [i] - v_{\rm obs,lead} [i])^2}{\sigma_{v}^2} \nonumber \\
 +  \frac{1}{n_{v,{\rm trail}} - 3} \sum_i \frac{(v_{\rm orbit} [i] - v_{\rm obs,trail} [i])^2}{\sigma_{v}^2} \nonumber \\
 +  \frac{1}{n_{\delta} - 3} \sum_i \frac{(\delta_{\rm orbit}[i] - \delta_{\rm obs}[i])^2}{\sigma_{\delta}^2}
\end{eqnarray}
where $n_{v,{\rm lead}} = 94$, $n_{v,{\rm trail}} = 108$, and $n_{\delta} = 17$, the number of observational data points in the 2MASS leading/trailing radial velocity samples
and the SDSS survey fields along the main branch of the Sgr stream respectively.
We adopt an uncertainty $\sigma_v = 12$\kms for each radial velocity measurement (this represents a combination of observational uncertainty and intrinsic stream width; 
see discussion in Majewski et al. 2004), and $\sigma_{\delta} = 1.9^{\circ}$ for each survey field location from Belokurov et al. (2006)\footnote{This uncertainty is a rough estimate 
of the uncertainty in the centroid of the stream at a given longitude based on the apparent width of the stream.}.
The radial velocity $v_{\rm orbit}$ and angular location $\delta_{\rm orbit}$ of the orbit at each longitudinal position ($\Lambda [i]$ or $\alpha [i]$ respectively) 
is determined via linear interpolation of the orbital path.

\begin{figure*}
\plotone{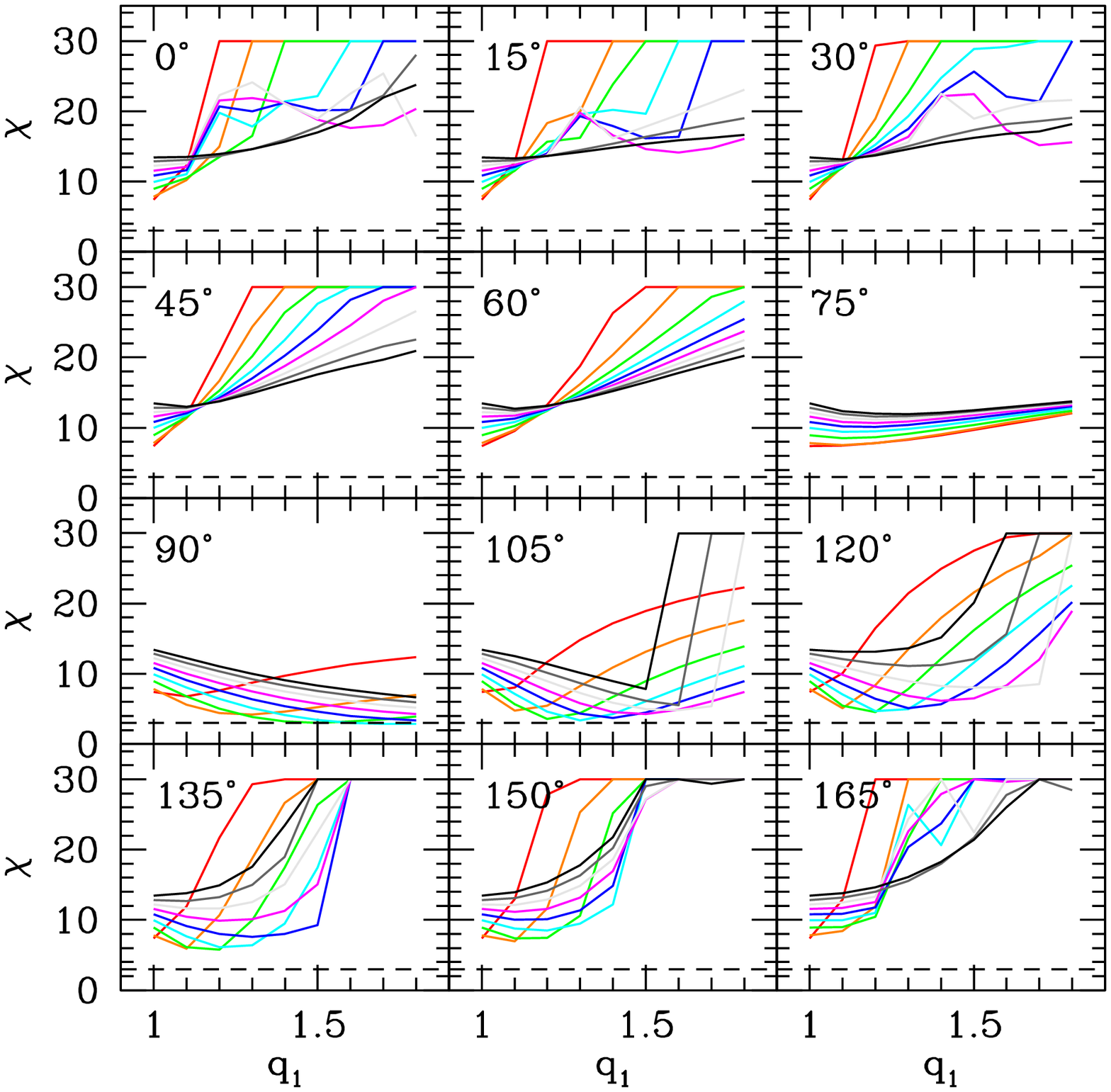}
\caption{Quality-of-fit parameter $\chi$ is shown as a function of the axial flattening $q_1$ in the Galactic Plane
for various choices of the axial rotation angle ($\phi$) and the flattening  perpendicular to the Galactic Plane ($q_z$).  Numbers in each panel represent the value of $\phi$, while
red/orange/green/cyan/blue/magenta/light grey/dark grey/black curves respectively correspond to $q_z = 1.0/1.1/1.2/1.3/1.4/1.5/1.6/1.7/1.8$.  The dashed line represents $\chi = 3.0$,
curves approach this line only for $\phi \sim 90^{\circ} - 105^{\circ}$.  Note that $\phi = 0^{\circ}$ corresponds to $q_1$ lying 
along the Galactic $X$ axis, and $\phi = 90^{\circ}$ to the Galactic $Y$ axis.}
\label{phifig.fig}
\end{figure*}


In Figure \ref{phifig.fig}  we show the values of $\chi$ resulting from various choices for $\phi$, $q_1$, and $q_z$.  
Values of $\phi \approx 90^{\circ} - 105^{\circ}$ (i.e. $q_1$ nearly aligned with the Galactic $Y$ axis) are strongly favored and are the only
models in which values of $\chi < 5$ can be obtained.
Given the limitations of the orbit-fitting method, it is not possible to conclusively discriminate between $q_1$--$q_z$ pairs in the triaxial region of parameter space
$q_z \approx 1.2 - 1.4$ (green/cyan/blue lines respectively in Fig. \ref{phifig.fig}) and $q_1 \approx 1.2 - 1.8$
since deviations of the actual tidal debris from the orbital path of the bound satellite are expected to have an effect comparable in magnitude to the computed values of $\chi \approx 3$.
More detailed fitting incorporating finer sampling scales in $q_z$, $q_1$, and $\phi$
in combination with comprehensive N-body models is beyond the scope of the present work and will be presented by
Law et al. (in prep); at present we note simply that the absolute minimum ($\chi = 2.7$) occurs for
$\phi = 90^{\circ}$, $q_z = 1.25$, $q_1 = q_y = 1.5$.
The Laplacian of this potential is everywhere positive, indicating that it corresponds to a physically realizable density distribution.



\section{DISCUSSION}
\label{discussion.sec}

In LJM05 we conducted an extensive parameter space search exploring the effects of  varying the Galactic disk/bulge mass,
the dark halo scale length, Sgr kinematics, the distance to the Galactic Center, the
distance to Sgr, and the total mass scale of the Milky Way.  No variation of these parameters gave rise to a model in which it was possible to simultaneously reproduce all three observational
constraints (trailing velocities, leading precession, leading velocities) on the Sgr stream in an axisymmetric Galactic halo.
In Figure \ref{ObsData.fig} we overplot on the observational constraints the orbits in an axisymmetric halo that best fit individual constraints:
the black curve is chosen to best-fit the angular precession of the leading arm ($q_1 = 1.00$, $q_z = 0.97$), 
while the green curve is chosen to best fit the radial velocity trend
of the leading arm ($q_1 = 1.00$, $q_z = 1.25$).  These models have quality-of-fit parameters
$\chi = 8.3/9.4$ respectively.  In contrast, the red curve in Figure \ref{ObsData.fig} represents our best-fit orbit in a triaxial halo with $q_z =1.25$, $q_1 = 1.50$, $\phi = 90^{\circ}$: 
in this model the orbit of the Sgr dwarf
simultaneously reproduces both the observed run of angular precession and radial velocity for the leading arm.
The corresponding ratios for the minor/major ($c/a = 0.67$) and intermediate/major 
($b/a = 0.83$) axis ratios in this triaxial halo are broadly consistent with the typical values predicted by cosmological simulations (e.g., Hayashi et al. 2007),
although the best-fit triaxiality parameter $T \approx 0.56$ is somewhat lower than anticipated.

While the fit may change somewhat with detailed N-body models, we can conclude 
not only that the Galactic dark halo may be triaxial within the orbit of Sgr ($r \lesssim 60$ kpc), 
but that the short/long axes of the halo may be aligned with the Galactic
$X/Y$ axes respectively to within $\sim 15^{\circ}$.
If the axial flattenings are introduced into an NFW model of the dark matter density
distribution, qualitatively similar results are obtained with $q_{\rho, x}/q_{\rho, y}/q_{\rho, z} = 1.0/2.0/1.66$.  As expected (e.g., Kuhlen et al. 2007) the isovelocity contours
are more spherical than the isodensity contours, but both indicate a similar preference for a triaxial system in which the major/minor axes are approximately aligned
with the Galactic $Y$/$X$ axes respectively.

This alignment is somewhat similar to that of the {\it stellar} halo described by Newberg \& Yanny (2006), who found a major axis similarly aligned with the Galactic $Y$ axis to
within $\sim 20-40^{\circ}$, albeit with a minor axis along $Z$ rather than $X$.  There is no clear correlation however
with the triaxiality of the Galactic bulge: the long axis
of the bar is currently though to lie within $\sim 15-20^{\circ}$ of the Galactic $X$ axis (e.g., Nakada et al. 1991; Morris \& Serabyn 1996; Babusiaux \& Gilmore 2005),
close to the minor axis of the dark halo.

In closing, we note two complications of the above results.
First, our analysis has assumed that the disk lies in one of the symmetry planes of the halo potential and that the axes themselves do not twist with radius. 
While such assumptions should be explored with future studies, we do not expect the broad results of our study to change --- 
an orientation significantly out of a symmetry plane would result in warping of the disk, which would dampen the misalignment on
timescales of a few disk orbits (e.g., Dubinski \& Kuijken 1995; Bailin et al. 2007; Jeon et al. 2009).
Second,  in the family of models presented here the Galactic $Z$ axis is the intermediate axis of the triaxial ellipsoid.  In a global sense such models
are somewhat unsatisfactory since  orbits about this axis (i.e. in the plane of the Galactic disk) are expected to be unstable (e.g., Binney et al. 1981).  At present, however,  
such models are the only ones
in which it is possible to simultaneously reproduce all of the observed characteristics of the Sgr stream.


\acknowledgements

Support for this work was provided by NASA through Hubble Fellowship grant \# HF-01221.01
awarded by the Space Telescope Science Institute, which is operated by the Association of Universities for Research in Astronomy, Inc., for NASA, under contract NAS 5-26555.
SRM acknowledges support from National Science Foundation grant AST-0807945.

\end{document}